\documentclass[aps,prl,twocolumn,superscriptaddress,nobibnotes,showpacs]{revtex4}
\usepackage{graphicx}

\begin{document}


\title{Prompt dipole radiation in fusion reactions}


\author{B.~Martin}
\affiliation{Dipartimento di Scienze Fisiche, Universit\`{a} di
Napoli "Federico II", via Cintia, 80125 Napoli, Italy}
 \affiliation{INFN,
Sezione di Napoli, via Cintia, 80125 Napoli, Italy}
\author{D.~Pierroutsakou}
\affiliation{INFN, Sezione di Napoli, via Cintia, 80125 Napoli,
Italy}

 \author{C. Agodi}
 \affiliation{INFN, Laboratori Nazionali del Sud, via S. Sofia, 95123 Catania, Italy}
        \author{R. Alba}
        \affiliation{INFN, Laboratori Nazionali del Sud, via S. Sofia, 95123 Catania, Italy}
    \author{V. Baran}
            \affiliation{INFN, Laboratori Nazionali del Sud, via S. Sofia, 95123 Catania,
            Italy},
              \affiliation{University of Bucharest, Romania}
               \affiliation{NIPNE-HH, 077125 Magurele, Romania}
        \author{A.~Boiano}
\affiliation{INFN, Sezione di Napoli, via Cintia, 80125 Napoli,
Italy}
        \author{G. Cardella}
        \affiliation{INFN, Sezione di Catania, 95123 Catania,
        Italy}
        \author{M. Colonna}
        \affiliation{INFN, Laboratori Nazionali del Sud, via S. Sofia, 95123 Catania,
        Italy},
         \affiliation{Dipartimento di Fisica, Universit\`{a} di Catania, 95123 Catania,
        Italy}
        \author{R. Coniglione}
        \affiliation{INFN, Laboratori Nazionali del Sud, via S. Sofia, 95123 Catania, Italy}
        \author{E. De Filippo}
        \affiliation{INFN, Sezione di Catania, 95123 Catania,
        Italy}
        \author{A. Del Zoppo}
        \affiliation{INFN, Laboratori Nazionali del Sud, via S. Sofia, 95123 Catania, Italy}
        \author{M. Di Toro}
        \affiliation{INFN, Laboratori Nazionali del Sud, via S. Sofia, 95123 Catania, Italy}
         \affiliation{Dipartimento di Fisica, Universit\`{a} di Catania, 95123 Catania,
        Italy}
        \author{G.~Inglima}
\affiliation{Dipartimento di Scienze Fisiche, Universit\`{a} di
Napoli "Federico II", via Cintia, 80125 Napoli, Italy},
 \affiliation {INFN,
Sezione di Napoli, via Cintia, 80125 Napoli, Italy}
        \author{T. Glodariu}
        \affiliation{NIPNE-HH, 077125 Magurele, Romania},
\author{M.~La Commara}
\affiliation{Dipartimento di Scienze Fisiche, Universit\`{a} di
Napoli "Federico II", via Cintia, 80125 Napoli, Italy},
 \affiliation{INFN, Sezione di Napoli, via Cintia, 80125 Napoli, Italy}

        \author{C. Maiolino}
        \affiliation{INFN, Laboratori Nazionali del Sud, via S. Sofia, 95123 Catania, Italy}
        \author{M. Mazzocco}
        \affiliation{Dipartimento di Fisica and INFN, Sezione di Padova, 35131 Padova, Italy}
        \author{A. Pagano}
        \affiliation{INFN, Sezione di Catania, 95123 Catania,
        Italy}
        \author{P. Piattelli}
        \affiliation{INFN, Laboratori Nazionali del Sud, via S. Sofia, 95123 Catania, Italy}
        \author{S. Pirrone}
        \affiliation{INFN, Sezione di Catania, 95123 Catania,
        Italy}
        \author{C. Rizzo}
        \affiliation{INFN, Laboratori Nazionali del Sud, via S. Sofia, 95123 Catania,
        Italy},
         \affiliation{Dipartimento di Fisica, Universit\`{a} di Catania, 95123 Catania,
        Italy}
\author{M.~Romoli}
\affiliation{INFN, Sezione di Napoli, via Cintia, 80125 Napoli,
Italy}
\author{M.~Sandoli}
\affiliation{Dipartimento di Scienze Fisiche, Universit\`{a} di
Napoli "Federico II", via Cintia, 80125 Napoli, Italy},
 \affiliation{INFN, Sezione di Napoli, via Cintia, 80125 Napoli, Italy}
        \author{D. Santonocito}
        \affiliation{INFN, Laboratori Nazionali del Sud, via S. Sofia, 95123 Catania, Italy}
        \author{P. Sapienza}
        \affiliation{INFN, Laboratori Nazionali del Sud, via S. Sofia, 95123 Catania, Italy}
        \author{C. Signorini}
        \affiliation{Dipartimento di Fisica and INFN, Sezione di Padova, 35131 Padova, Italy}

\date{\today}

\begin{abstract}
The prompt $\gamma$-ray emission was investigated in the 16A MeV
energy region by means of the $^{36,40}$Ar+$^{96,92}$Zr fusion
reactions leading to a compound nucleus in the vicinity of
$^{132}$Ce. We show that the prompt $\gamma$ radiation, which
appears to be still effective at such a high beam energy, has an
angular distribution pattern consistent with a dipole oscillation
along the symmetry axis of the dinuclear system. The data are
compared with calculations based on a collective bremsstrahlung
analysis of the reaction dynamics.

\end{abstract}



\maketitle

 \vspace{-4cm}

 The study of the Giant Dipole Resonance
(GDR) decay from excited nuclei is a topic of central importance in
nuclear physics because it constitutes a powerful probe to get
insight into the features of nuclei far from normal conditions. It
was suggested many years ago in \cite{berl79} that the charge
equilibration mechanism occurring in dissipative heavy-ion
collisions could be related to the direct excitation of a GDR in the
composite system. Subsequently, this idea was considered within
various theoretical frameworks
(\cite{refchom93,refpapa,refsim2001,refdasso,refbaran_npa},
\cite{refbaran_prl} and references therein) leading to similar
conclusions: at the very early stages of charge-asymmetric heavy-ion
collisions a large amplitude collective dipole oscillation, the
so-called dynamical dipole mode, can be triggered along the symmetry
axis of the strongly deformed composite system. This oscillation
could decay emitting prompt dipole photons, in addition to the
photons originating from the GDR thermally excited in the hot
compound nucleus (CN), with: i)a lower energy than that of a GDR
built in a spherical nucleus of similar mass and ii)an anisotropic
angular distribution. Different parameters are predicted to
influence the prompt dipole photon intensity: besides the entrance
channel charge asymmetry also the mass asymmetry \cite{refbaran_prl}
and the incident energy \cite{refchom93,refsim2001, refbaran_prl}
should play a role.

From an experimental point of view there have been attempts
\cite{ref1, ref2,ref3,ref4,ref5,prc} to evidence this
pre-equilibrium emission by using the following technique: formation
of the same composite system from entrance channels with different
charge asymmetry. The comparison of the associated $\gamma$-ray
spectra in the above measurements evidenced an excess in the
composite system GDR energy region for the more charge asymmetric
system that was identified with the prompt dipole radiation.
However, to date no experimental information exists on the angular
distribution of this $\gamma$-ray excess that allows to draw firm
conclusions on its origin. Furthermore, there is no systematic study
of the phenomenon as a function of the reaction parameters that are
predicted to influence it.

In our previous works \cite{ref5, prc} we started an investigation
of the prompt dipole radiation for the $^{32,36}$S+$^{100,96}$Mo
reaction pair at incident energies E$_{lab}$=6A and 9A MeV. In these
measurements, a CN in the Ce mass region was formed with excitation
energy E$^{*}$=117 and 174 MeV, respectively. In the present letter
we extend our investigation by using the $^{36}$Ar+$^{96}$Zr and
$^{40}$Ar+$^{92}$Zr reactions at E$_{lab}$=16 and 15.1A MeV,
respectively, to form a CN in the same mass region (average mass
A$\sim$126) at an average excitation energy E$^{*}$$\sim$280 MeV. We
present the first experimental evidence that the prompt radiation is
related to a dipole oscillation along the dinuclear system symmetry
axis, by studying its angular distribution.
 We infer the
  characteristics of this oscillation and their dependence on beam energy.
 Finally, we compare our data with calculations based on a
collective bremsstrahlung analysis of the reaction dynamics.

\begin{table*}[hbt]
\caption{\it Reaction pair, incident energy, compound nucleus
excitation energy, initial dipole moment D(t=0), initial mass
asymmetry $\Delta$, percent increase of the intensity in the
linearized $\gamma $-ray spectra for the more charge asymmetric
system (the energy integration limits, in MeV, are given in the
parenthesis), centroid energy E$_{dd}$ and width $\Gamma$$_{dd}$ of
the dynamical dipole mode obtained by the fit of the data described
in the text.}

\begin{ruledtabular}
\begin{tabular}{lcccccccccc}
Reaction  & E$_{lab}$(MeV/nucleon)& E$^{*}$(MeV)& D(t=0) (fm) & $\Delta$ &  Increase (\%)  &  E$_{dd}$ (MeV)  &$\Gamma$$_{dd}$ (MeV) &Ref \\
\hline
$^{32}$S+$^{100}$Mo  & 6.125 & 117 &18.2 & 0.19 &  1.6 $\pm 2.0$ (8,21) & & &  \cite{prc}\\
$^{36}$S+$^{96}$Mo & 5.95 & 117 &1.7 & 0.16 &  \\
\hline
$^{32}$S+$^{100}$Mo & 9.3& 174 &18.2 & 0.19 & 25 $\pm 2$ (8,21) & 11.4$\pm0.3$ &3.0$\pm0.5$ &\cite{ref5}\\
$^{36}$S+$^{96}$Mo &  8.9& 174 &1.7 & 0.16  \\
\hline
$^{36}$Ar+$^{96}$Zr & 16 & $\sim$280 & 20.6 & 0.16 & 12 $\pm 2$  (8,21)& 12.2$\pm0.6$ & 3.7$\pm1.4$  & present data \\
$^{40}$Ar+$^{92}$Zr & 15.1 & $\sim$280 &4.0 & 0.14 &  \\

\end{tabular}
\end{ruledtabular}
\end{table*}

All the studied reaction pairs leading to compound nuclei in the Ce
mass region are presented in Table I, along with the entrance
channel relevant quantities, i.e. the incident energy, the CN
excitation energy, the initial dipole moment D(t=0) and the mass
asymmetry $\Delta$ (for a definition of the dipole moment and mass
asymmetry see \cite{prc}). The technique used in the present work is
the same described in \cite{ref5,prc}, that is all reaction
parameters were identical for the two systems apart from the initial
dipole moment. From Table I we can see that the dipole moment
changed by 16.6 fm from the $^{40}$Ar~+~$^{92}$Zr system to the more
N/Z asymmetric one, $^{36}$Ar~+~$^{96}$Zr, while the entrance
channel mass asymmetry changed by a very small amount. The critical
angular momentum for fusion events was equal for both reactions,
according to PACE2 calculations \cite{ref9}, avoiding thus any
difference in the CN spin distribution. Moreover, the CN excitation
energy was the same within errors in both reactions as it will be
shown later in the text. The results concerning the
$^{36,40}$Ar~+~$^{96,92}$Zr pair can be directly compared with those
related to the $^{32,36}$S+$^{100,96}$Mo one because of the similar
difference in the entrance channel dipole moment and mass asymmetry.

The $^{36,40}$Ar pulsed beams, provided by the Superconducting
Cyclotron of the Laboratori Nazionali del Sud (LNS), impinged on a
450 $\mu $g/cm$^{2}$-thick $^{96}$ZrO$_{2}$ and on a 600 $\mu
$g/cm$^{2}$-thick $^{92}$ZrO$_{2}$ target enriched to 95.63$\%$ in
$^{96}$Zr and to 95.36$\%$ in $^{92}$Zr, respectively. The targets
were evaporated on carbon layers 90 and 60 $\mu $g/cm$^{2}$-thick,
respectively. The $\gamma$-rays (E$_\gamma$$>$5.5 MeV) and the light
charged particles were detected by using the 180 BaF$_{2}$ modules
of the MEDEA experimental apparatus \cite{ref12} that covers the
polar angular range between $\theta$=30$^{\circ }$ and
$\theta$=170$^{\circ}$
 and the full range in the azimuthal angle $\phi$. The fusion-evaporation residues were detected by four Parallel Plate
Avalanche Counters (PPAC's) located symmetrically around the beam
direction at 70 cm from the target, centered at
$\theta$=7${{}^{\circ }}$ and covering 7${{}^{\circ }}$ in $\theta$.
They provided the time of flight with respect to the radiofrequency
signal of the accelerator and the energy loss of the reaction
products. Down-scaled single events together with coincidence events
between a PPAC and at least one fired BaF$_{2}$ scintillator were
collected during the experiment. The energy calibration of the
$\gamma$-ray detectors was obtained by using the composite
radioactive sources of $^{241}$Am+$^{9}$Be and $^{238}$Pu+$^{13}$C
and the 15.1 MeV $\gamma$-rays from the p+$^{12}$C reaction while
the calibration of the charged particles was performed as described
elsewhere \cite{calib}. The discrimination between $\gamma$-rays,
light charged particles and neutrons was performed by combining a
pulse shape analysis of the BaF$_{2}$ signal and a time of flight
measurement with respect to the radiofrequency signal of the
Cyclotron.

At the present incident energies, the incomplete fusion cross
section represents approximately the 90$\%$ of the total fusion
cross section \cite{ref13}. The average excitation energy and the
average mass carried away by the pre-equilibrium particles was
evaluated by analyzing the energy spectra of the protons and alpha
particles detected at different angles in coincidence with the
evaporation residues. The particle spectra were simultaneously
fitted in the hypothesis of two moving sources (see for example
\cite{santo02}). A slow source having the center-of-mass velocity
which simulates the statistical particle emission from the CN and an
intermediate-velocity source that represents the emission of fast
particles of non statistical origin. Details on this analysis will
be presented elsewhere. The relevant information extracted from the
above work concerns the CN average excitation energy and average
mass that was found to be identical within errors for the two
reactions. This allows us to compare safely the associated
$\gamma$-ray spectra with each other. The CN excitation energy
estimated from our data (see Table I) is somewhat lower than that
predicted by the empirical formula given in \cite{ref15} for the
$^{18}$O+$^{100}$Mo system, according to which, the excitation
energy in the present case is expected to be E$^{*}$$\sim$300 MeV.

In Fig.1a we present the bremsstrahlung-subtracted $\gamma$-ray
spectra of the $^{40}$Ar+$^{92}$Zr (open circles) and
$^{36}$A+$^{96}$Zr (filled squares) reaction, taken at
$\theta$=90$^{\circ}$ in coincidence with the evaporation residues
and integrated over 4$\pi$ assuming an isotropic angular
distribution. $\epsilon$$_{det}$
 is the energy dependent efficiency of the
experimental apparatus. The bremsstrahlung component was deduced by
fitting simultaneously the $\gamma$-ray spectra of both reactions
for E$_{\gamma}$~$\geq$~35 MeV at different angles by means of an
exponential function with isotropic emission in a reference frame
moving with 0.5$\textit{v}$$_{beam}$ \cite{refnif}. The difference
between these spectra, displayed in the same figure with the stars,
shows an excess of $\gamma$-rays emitted during the charge
asymmetric reaction ($^{36}$Ar+$^{96}$Zr) and concentrated at
E$_{\gamma}$$\sim$12 MeV, that is in the energy region of the CN
GDR. This excess is related to entrance channel charge asymmetry
effects, being identical all the other reaction parameters and it is
attributed to the dynamical dipole mode present at the beginning of
the dinuclear system formation. To deduce the centroid energy
E$_{dd}$ and the width $\Gamma$$_{dd}$ of the dynamical dipole mode,
the observed $\gamma$-ray excess was fitted
 by means of a lorentzian curve folded by the
experimental apparatus response function (solid line of Fig. 1a)
\cite{reffoldmedea}. The parameters extracted from the fit are seen
in Table I for both the present data and the data taken at 9A MeV.
For both beam energies, E$_{dd}$ was found to be lower than the
centroid energy of the CN GDR (E$_{GDR}$=14 MeV), implying a
deformation of the composite system at the moment of the prompt
dipole $\gamma$-ray emission. In a naive picture of two colliding
nuclei at
the touching configuration, we expect E$_{dd}$$\sim\frac{%
78}{A^{1/3}_{1}+A^{1/3}_{2}} $$\sim$ 10 MeV, $A_{1}$ and $A_{2}$
being the colliding ion masses. The fact that it was found to be
somewhat larger than that expected, nicely confirms that some
density overlap already exists at the start up of the dipole
oscillation \cite{refbaran_npa}. It is interesting to notice from
Table I that centroid energy and width remain constant within errors
by increasing the beam energy.

The details in the GDR energy region can be better evidenced if the
$\gamma$-ray spectra of Fig. 1a are linearized, dividing them by the
same theoretical spectrum. This theoretical spectrum was obtained by
using the statistical decay code CASCADE \cite{refpuh} with a
constant dipole strength function and folded by the response
function of the experimental apparatus. The resulting linearized
data are shown with the same symbols in Fig. 1b. The solid line in
the figure depicts the linearized theoretical $\gamma$-ray spectrum
of the charge symmetric reaction calculated with the CASCADE code
and folded by the experimental setup response function. For the
calculation, the following parameters were used: a CN mass A=126,
E$^{*}$=284 MeV and a level density parameter which varies with
nuclear temperature according to (\cite{pier96} and ref. 26
therein). The GDR strength function was taken to be a lorentzian
curve with centroid energy E$_{GDR}$=14 MeV, width $\Gamma _{GDR}$ =
13 MeV and strength S$_{GDR}$=100$\%$ of the E1 energy-weighted
sum-rule strength. Moreover, a cutoff in the $\gamma$-ray emission
for excitation energies larger than E$^{*}$=250 MeV was applied, in
good agreement with \cite{suom96} for nuclei in the A$\sim$ 115 mass
region.

 If the linearized data of Fig. 1b are integrated between 8 and 21 MeV, an increase
of the $\gamma$-ray intensity of 12\% is found in the more charge
asymmetric system. From Table I, where the percent increase of the
linearized spectra for the three beam energies is shown, we can see
that the prompt dipole radiation intensity presents a maximum at 9A
MeV decreasing toward lower and higher energies. Although diminished
with respect to its value at 9A MeV, it is still observed at nuclear
excitation energies as high as $\sim$280 MeV, excluding a fast
increase of the dynamical dipole mode damping width with excitation
energy. In fact the dynamical dipole mode is a pre-equilibrium
collective oscillation present before the thermalization of the
mechanical energy. The damping is then also related to fast
processes, the pre-equilibrium nucleon emissions (mostly neutrons,
that are reducing the asymmetry) and (p,n) direct collisions that
will damp the isovector oscillation. From calculations we expect
that both mechanisms are smoothly increasing in the present range of
beam energies.

\vspace{-0.8cm}
\begin{figure}[htbp]%
\centering%
\hspace{-15mm}
\includegraphics[width=8cm,height=10cm]{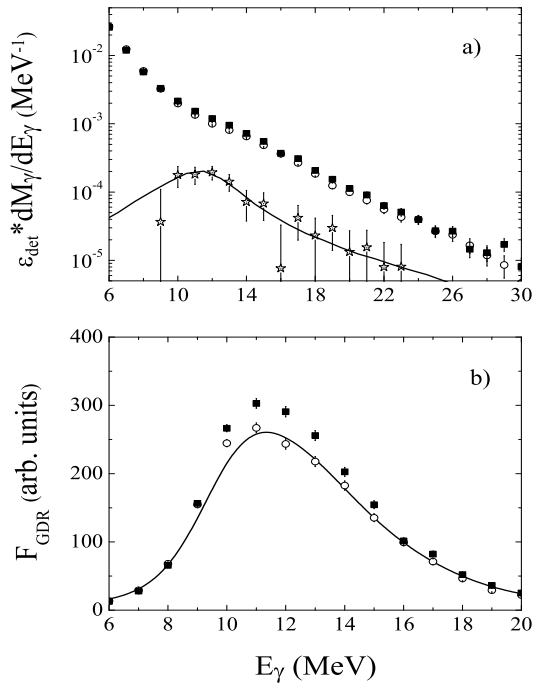}%

 \vspace{-0.5cm}
\caption{a)90$^{\circ}$ bremsstrahlung-subtracted $\gamma$-ray
spectrum of the
      $^{36}$Ar+$^{96}$Zr (filled squares) and the $^{40}$Ar+$^{92}$Zr (open circles) reaction in coincidence with evaporation residues
      along with
       their difference (stars). The solid line is the result of the fit
described in the text.
       b)90$^{\circ}$ bremsstrahlung-subtracted linearized
$\gamma$-ray spectra. The solid line represents the linearized
theoretical spectrum calculated with the code CASCADE for the charge symmetric reaction.}%
\end{figure}%

 \vspace{-0.3cm}

Up to now, experimental evidences of the dipole character of the
prompt $\gamma$-ray emission have never been reported in the
literature. To infer it, we display in Fig. 2 the center-of-mass
angular distribution with respect to the beam direction of the
observed $\gamma$-ray excess, integrated over energy from 10 to 15
MeV. The lines in the figure depict the expected angular
distribution given by the Legendre polynomial expansion:
$W(\theta_{\gamma} )=W_{0}[1+a_{2}P_{2}(cos(\theta_{\gamma})]$ for
different values of the anisotropy coefficient $a_{2}$. In all
cases, the coefficient $W_{0}$  was obtained from a best fit to the
 data. When
$a_{2}=-1$ the angular distribution takes the
$sin^{2}(\theta_{\gamma})$ form of pure dipole emission (solid
line), while $a_{2}=-0.8$ (dashed line) and $a_{2}=-0.5$ (dotted
line) correspond to more diffuse angular distributions.
  We see that the experimental angular distribution is strongly anisotropic with a maximum around
 90$^{\circ}$, consistent with emission from a dipole oscillating along the
  beam axis (solid line). For
near-central collisions as in the present case, the symmetry axis of
the dinuclear composite system is nearly coincident with the beam
axis at the very early moments of its formation. In the case of a
larger mean inclination of the axis of the direct dipole oscillation
because rotation has taken place meanwhile, we would expect a
widening of the angular distribution with respect to 90$^{\circ}$
(dashed and dotted lines). This effect should be directly related
to: a)the rotation angular velocity of the dinuclear system during
the prompt dipole emission b)the instant at which this emission
occurs. The data suggest that the oscillation axis of the direct
dipole has not rotated much with respect to the beam direction. This
result is compatible with emission of the prompt dipole radiation at
the very beginning of the reaction. In perspective, we can say that
accurate measurements of the dynamical dipole angular distribution
could even allow to directly evaluate the corresponding mean
rotation of the emitting dinuclear system and so the time scale of
such pre-equilibrium $\gamma$ radiation.

 \vspace{-0.6cm}
\begin{figure}[htbp]%
\centering%
\hspace{-10mm}
\includegraphics[width=9cm,height=7cm]{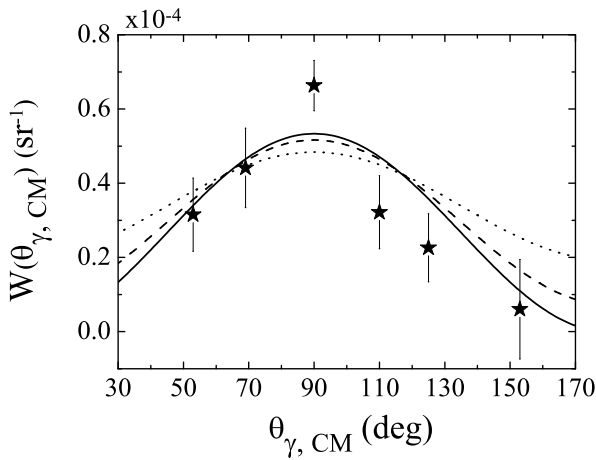}%

 \vspace{-0.7cm}
\caption{Center-of mass angular distribution of the difference
between the data of the two reactions for $\gamma$-rays with 10
MeV$\le$E$_{\gamma}$$\le$15 MeV. The lines are described in the text.}%
\end{figure}%

 \vspace{-0.3cm}

 Calculations of the prompt dipole radiation for the
 $^{36}$Ar+$^{96}$Zr and
 $^{32}$S+$^{100}$Mo reactions at 16A MeV were performed within the BNV transport model framework
and based on a collective bremsstrahlung approach
\cite{refbaran_prl,prc, rizzo07}. In the transport calculations no
free parameters were used. The results for the above reactions are
identical within a 20\% uncertainty, justifying thus their direct
comparison with each other.
 For near-central collisions the total multiplicity of the prompt
dipole radiation was found to be $ 0.8$$\cdot$$10$$^{-3}$ and $
1.4$$\cdot$$10$$^{-3}$, depending on the NN cross section used. The
lower estimation is related to free NN cross sections, while the
upper one is obtained with the in medium reduced ones \cite{Li94}.
Reduced cross sections are leading to larger dipole radiation rates
for two reasons: i) less fast nucleons emission, in particular for
neutrons which directly decrease the dipole strenght; ii) reduced
attenuation of the dipole p-n oscillation due to a smaller number of
p-n direct collisions. The theoretical total multiplicity is in very
good agreement with the experimental one, ($ 0.7$ $\pm
0.1$)$\cdot$$10$$^{-3}$, obtained integrating the $\gamma$-ray
excess over energy and over solid angle by taking into account its
angular distribution and the experimental set up efficiency.
Moreover, the theoretical dynamical dipole centroid energy and
width, $E_{dd, th}$$\sim$ 9 MeV and $\Gamma_{dd,th}$$\sim$2 MeV, are
in reasonable agreement with the corresponding experimental values
(Table I).

 In summary, we present a study of the prompt dipole radiation in the
16A MeV energy region and we compare the present results with
previous ones obtained at lower incident energy. The outstanding
feature of our data is the angular distribution pattern of the
observed $\gamma$-ray excess which is consistent with that of a
dipole oscillating along the beam axis. This result suggests that
the prompt $\gamma$-ray emission occurs during the first stages of
the dinuclear system formation. Calculations based on a collective
bremsstrahlung analysis of the reaction dynamics predict
characteristics of the dynamical dipole mode that are in good
consistency with experiment.

We warmly thank M. Loriggiola and
 A. Stefanini (LNL, Italy) for providing the Zr targets and the LNS staff for
the excellent quality of the Ar beams.

\end{document}